\documentclass[12pt]{article}
\pdfoutput = 1
\usepackage{graphicx}
\usepackage{adjustbox}
\usepackage{color,soul}
\usepackage{fancybox}
\usepackage{colortbl}
\usepackage{dcolumn}
\usepackage{amsfonts,amsmath,amssymb}
\usepackage{appendix}
\usepackage{tikz} 
\usetikzlibrary{fit,positioning}
\usepackage[section]{placeins}
\usepackage{lineno}
\usepackage{subcaption}
\usepackage{float}
\usepackage{hyperref}
\usepackage{authblk}
\usepackage{array}
\usepackage{multirow,multicol}
\usepackage{nameref,hyperref}
\usepackage[superscript]{cite}

\providecommand{\keywords}[1]
{
  \small	
  \textbf{\textit{Keywords:}} #1
}

\usepackage{tabularx}
\newcolumntype{L}[1]{>{\raggedright\arraybackslash}m{#1}}
\newcolumntype{C}[1]{>{\centering\arraybackslash}m{#1}}
\newcolumntype{R}[1]{>{\raggedleft\arraybackslash}m{#1}}

\usepackage{verbatim}
\usepackage{moreverb} 

\title{A multilayer network model of Covid-19: implications in public health policy in Costa Rica}

\author[1, +]{Fabio Sanchez, PhD}
\author[1, +]{Juan G. Calvo, PhD}
\author[2, +]{Gustavo Mery, PhD}
\author[3,4, +]{Yury E. Garc\'ia, PhD}
\author[3, +]{Paola V\'asquez, MD}
\author[1, +]{Luis A. Barboza, PhD}
\author[3, +]{María Dolores Pérez, MD}
\author[5, +]{Tania Rivas, MD}

\affil[1]{Universidad de Costa Rica, Centro de Investigaci\'on en Matem\'atica Pura y Aplicada - Escuela de Matem\'atica, San Jos\'e, Costa Rica}
\affil[2]{Pan America Health Organization, World Health Organization, San José 10102, Costa Rica.}
\affil[3]{Universidad de Costa Rica, Centro de Investigaci\'on en Matem\'atica Pura y Aplicada, San Jos\'e, Costa Rica}
\affil[4]{Department of Public Health Sciences, 
University of California Davis, CA, USA}
\affil[5]{Ministry of Health, San José 10102, Costa Rica}

\affil[*]{fabio.sanchez@ucr.ac.cr}
\affil[+]{these authors contributed equally to this work}

\date{}
\begin{document}
\maketitle

\begin{abstract}

Successful partnerships between researchers, experts, and public health authorities have been critical to navigate the challenges of the Covid-19 pandemic worldwide. In this collaboration, mathematical models have played a decisive role in informing public policy, with findings effectively translated into public health measures that have shaped the pandemic in Costa Rica. As a result of interdisciplinary and cross-institutional collaboration, we constructed a multilayer network model that incorporates a diverse contact structure for each individual. In July 2020, we used this model to test the effect of lifting restrictions on population mobility after a so-called “epidemiological fence” imposed to contain the country's first big wave of cases. Later, in August 2020, we used it to predict the effects of an open and close strategy (the Hammer and Dance). Scenarios constructed in July 2020 showed that lifting restrictions on population mobility after less than three weeks of epidemiological fence would produce a sharp increase in cases. Results from scenarios in August 2020 indicated that the Hammer and Dance strategy would only work with $50\%$ of the population adhering to mobility restrictions. The development, evolution, and applications of a multilayer network model of Covid-19 in Costa Rica has guided decision-makers to anticipate implementing sanitary measures and contributed to gain valuable time to increase hospital capacity. 


\end{abstract}

\keywords{Covid-19, Network Model, Public Health, Non-pharmaceutical interventions, Computational Model}

\section*{Introduction}

The global health crisis that began with the emergence of the SARS-CoV-2 virus has triggered unprecedented collaborations and research activity among the scientific community to better understand the transmission mechanisms of this novel pathogen, project the burden to health care systems, and evaluate the potential effects of the exceptional interventions that had to be implemented to slow down the spread of the virus.~\cite{haghani2020} In this effort, mathematical and statistical models have proven to be valuable tools in forecasting possible scenarios. These models can provide health authorities and decision-makers a scientifically-based tool to provide insight into the adoption of public health measures, such as restrictions on human mobility and guidance for resource allocation.~\cite{MathReview} Amidst this pandemic, a variety of traditional mathematical models have been developed. However, modeling teams have struggled in the effort to capture the transmission dynamics of Covid-19 in different parts of the world, given the {\it live} learning approach of the disease behavior.~\cite{COVID19Groups,MathReview} \\ 

{Some} highly specialized interdisciplinary research teams have developed sophisticated mathematical, statistical, and computational models using publicly available data.~\cite{COVID19Groups,Chang,LondonForecast} However, these experiments must be tailored specifically to the circumstances of each geographic location using each location's available information. Modelling efforts have also faced the challenges of rapidly adapting to the constant changes in social behavior based on the various restrictive measures taken worldwide. In this sense, quantitative methods incorporating a more realistic population structure and associated contact networks provide a more flexible tool to explicitly model heterogeneity in host contact patterns. However, this data is not available to modelers promptly.\\   

{As shown in this article,} Costa Rica has provided the necessary conditions for modelling teams to collaborate with health authorities and develop evidence-based tools to analyze and project possible epidemic scenarios. {This central american country of 5,163,038 inhabitants reported its first Covid-19 patient on March 6, 2020.~\mbox{\cite{minsaonline}}}. In the days to come ass more cases were confirmed, public health officials began to implement interventions to promote and facilitate physical distancing {in the seven provinces and 82 cantons the country is divided in}. During the first weeks, the government announced the cancellation of all massive events, the closing of all schools and universities, borders, and other social gatherings sites, the implementation of teleworking for the public and private sector, and restriction to vehicular circulation, {however by March 22,2020 the virus had been detected in all seven provinces and by August 12, 2020 all of the 82 cantons of the county had detected Covid-19 cases}. In these early stages of the pandemic, a group of interdisciplinary researchers, experts in mathematical modeling and other areas, and public health officials started gathering in an attempt to use the best available mathematical and statistical tools to inform health authorities in preparation for the inevitable surge in cases to come, with all its medical and social implications. Our research group started developing a network model with different contact layers of Covid-19 in Costa Rica and gained progressive practice to inform public health authorities.\\

In the months following March 2020, a reduction in cases resulted in the government lifting several containment measures to balance public health safety with the impact on the country's economy. At the same time, the Costa Rican Social Security Fund (CCSS), the country's main public health care provider, and other public and private institutions were working to increment hospital capacity gradually. Early in the pandemic, CCSS had at their disposal less than 100 intensive care units (ICU) beds and was in urgent need to increase ICU capacity and several other competing health priorities.\\

In July 2020, the Ministry of Health announced the inability to trace the source of infection of $65\%$ of the positive cases.~\cite{minsaonline} During that time, health authorities were working diligently to contact trace confirmed cases; this became increasingly difficult as the number of patients rapidly increased. The government's primary objective at the time was to prevent the collapse of the hospital care system. The country established a four-color alert status for every canton (green, yellow, orange, and red) to modify restrictions according to local infections.~\cite{minsaonline} On July 11-19, 2020, a reduction in mobility and closure of businesses was implemented in the cantons with orange alert status, located mainly in the Greater Metropolitan Area. This strategy was called "epidemiological fence" by the Costa Rican government.~\cite{minsaonline} During August 2020, the government implemented an open/close strategy called ``dance and hammer."~\cite{pueyo2020coronavirus} This strategy involved alternating between 9 days of less restriction in commercial activity and vehicular circulation in cantons with an orange alert and 12 days of closure in commercial activity and limiting vehicular circulation to two days a week for people living in these cantons.\\

{This article describes the multilayer, temporal and stochastic model developed for Costa Rica, tailored to project and inform decision-makers of probable transmission scenarios and the impact of the health care services under diverse public health interventions. To depict this collaboration, we present the simulations obtained using this model during July and August 2020, and analyzed its role in providing insights to the Costa Rican health authorities.}

\section*{Methods}

\subsection*{Network Model}
A network is a mathematical object composed of nodes and edges (links between vertices), that in our case, represent individuals in the population and interactions among them, respectively. In our model, the number of nodes in the graph is equal to the total population in Costa Rica. An edge connects two nodes on a specific day if they have contact. We try to mimic how the contacts of a node (individual) change over time by changing its number of edges (connections with neighbors) daily. The degree of a node refers to the number of edges at a particular time.\\

\begin{figure}[!ht]
\centering
\includegraphics[scale=0.4]{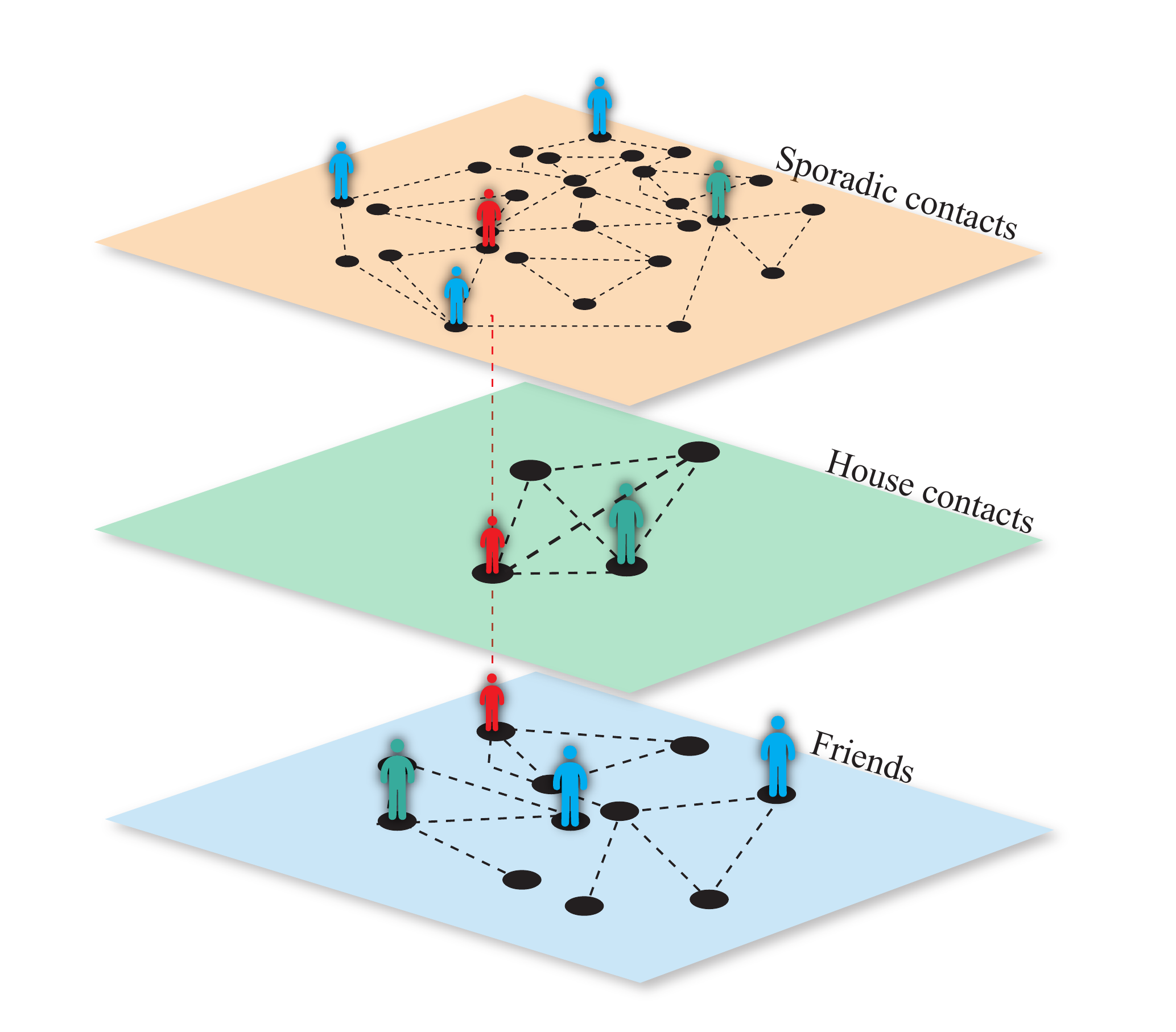}
\caption{{\bf Representation of contact layers in the network.} Three different layers are included in the network: household, social, and sporadic contacts.}
\label{fig:NetworkModel}
\end{figure}

To differentiate types of contacts in the population, we consider a temporal multilayer network; see.~\cite{porter2019nonlinearity} Moreover, we consider three layers as depicted in Fig~\ref{fig:NetworkModel}. {Each layer 
represents a different type of contact between individuals.} These layers are: (1) a {\it household network} (individuals that live in the same house), (2) a {\it social network} (known contacts such as friends and colleagues), and (3) a {\it sporadic network} (strangers that you may encounter in short periods when you visit random locations). Networks are connected by intra-layer edges (links in one layer) and inter-layer edges (links between layers). Layers one and two are fixed for each simulation, while layer three can change each day since an individual has no control over sporadic encounters.\\

{Each node has two types of attributes: (1) fixed parameters, such as canton of residence, household members, 
degree and graph connectivity for layer two, and (2) variable attributes, such as connectivity for layer three, epidemiological state, number of days at current epidemiological state, number of interactions with social contacts per day, use of personal protective measures, and self-care behavior.} The former is defined at the beginning of each simulation, and the latter can take random values every day accordingly. {These attributes such as canton of residence is used for connectivity between individuals; household members, interactions, degree and graph connectivity for the different layers, are used to construct the multilayer network. Epidemiological state, number of days at current epidemiological state, use of personal protective measures, and self-care behavior are used for the epidemiological model depicted in Figure \mbox{\ref{fig:EpiModel}}.} For implementation details, we refer to.~\cite{calvo2021multilayer}

\subsection*{SARS-CoV-2 transmission model} \label{EpidemicBehavior}

\begin{figure}[!ht]
\centering
\includegraphics[scale=0.05]{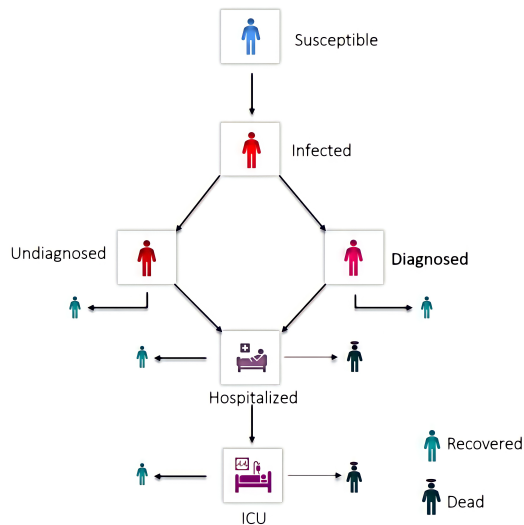}
\caption{{\bf Transmission model.} Individual's epidemiological states for SARS-CoV-2.}
\label{fig:EpiModel}
\end{figure}

The classical SEIR type dynamics is incorporated into the multilayer network model to describe the SARS-CoV-2 transmission process. The population is divided according to the epidemiological status. Thus, the first compartment corresponds to susceptible individuals. When susceptible individuals become infected, they move to a latency stage where eventually they become infectious (undiagnosed or diagnosed). Individuals could potentially recover, require a hospital bed, move into an ICU or die; see Fig~\ref{fig:EpiModel}. {We use a discrete time step (one day) for transitions between states.}\\

A susceptible individual can become infected by computing its probability of infection, which depends on its daily interactions with its infectious contacts (nodes). If the probability that node $j$ infects a susceptible node $i$ at a given day is $\beta_{ij}$, then the probability of infection $p_i$ for node $i$ is given by
\begin{eqnarray*}
p_i = 1-\prod_{j\neq i} (1-\beta_{ij}),
\end{eqnarray*}
where $j$ includes all the indices of nodes that can infect node $i$ that particular day. \\

The value $\beta_{ij}$ may depend on different attributes (such as layer, epidemiological state, human mobility, use of personal protective measures and self-care behavior, time of exposure, environment, and other factors). Depending on the interaction, the value of $\beta_{ij}$ can be reduced according to contact type (layer), personal protective measures, human mobility, and sneezing/coughing protocols, among others. In this way, we mimic the different health measures by setting
\begin{eqnarray}
\label{eq:pi2}
p_i = 1-\prod_{j\neq i} (1-c_j \beta_{ij})^{\gamma}
\end{eqnarray}
where $c_j$ is a reduction percentage that depends on the layer, use of personal protective measures, and human mobility in the interaction between nodes $i$ and $j$, respectively {(see Table \mbox{\ref{table1}})}, and $\gamma \in (0,1]$ is a calibration parameter estimated for each simulation. {For approximating $\gamma$, we consider a period of seven days where we calibrate the model by using a standard bisection method for finding an optimum value for $\gamma$, such that the square error between observed and predicted cases is minimum. The value of $\gamma$ is then fixed for all the simulations.} 


\subsection*{Data}

We describe each set of data that we used in our analysis as follows:

\begin{itemize}
\item \textit{Contact tracing}. The initial conditions of the model are based on the contact tracing provided by the Ministry of Health.~\cite{minsaonline} This data allowed us to create a more realistic contact network for the early stages of the pandemic. However, starting in July 2020, the number of cases increased, and contact tracing became a problem for public health authorities. From July 2020, as community transmission was declared, we only incorporate the number of daily cases per canton provided by the Ministry of Health.~\cite{minsaonline}

\item \textit{Household layer}. The number of households per canton and the average number of individuals per household was taken from the ``National Household Survey'',~\cite{Encuesta69:online} made by the National Institute of Statistics and Census (INEC) in 2018. In our model, the number of inhabitants of each household is determined by a Poisson distribution with a mean equal to the average number of inhabitants per household per canton. Total connectivity is assumed in each household with no reduction in the probability of infection ($c_j = 1$).

\item \textit{Mobility matrix}. Possible connections between nodes in different cantons are selected according to a connectivity matrix that provides the place where individuals reside and work, according to the ``National Household Survey".~\cite{Encuesta69:online} Connections (edges) between individuals of two different cantons may exist if there is mobility between them. The degree of a node in layers two and three is calculated using a uniform distribution on a given interval;~\cite{mossong2008social} see values in Table~\ref{table1}. We assume a fixed set of social contacts in layer two, but daily interactions are limited by choosing a subset of its edges.

\begin{table}[!ht]
\caption{{\bf Network parameters.} ({\it Left}) Degree node distribution for layers two and three; GMA stands for Greater Metropolitan Area. ({\it Right}) Infection parameters for Eq~\eqref{eq:pi2}. PM and {HM} stand for protective measures and human mobility, respectively.}
\centering
\begin{tabular} {|l|l|} \hline
 & Interval \\ \hline
Contacts in sporadic network& $[0,20]$\\ \hline
Contacts in social network & $[5,30]$  \\ \hline
Social contacts per day, GMA & $[5,25]$ \\ \hline
Social contacts per day, off GMA & $[5,15]$\\ \hline
\end{tabular}
\quad
\begin{tabular}{|c|c|c|} \hline
& $c_j$ & {Ref.} \\\hline
Sporadic contacts & 0.5000 & {Est.}\\ \hline
PM & 0.1476 & {\mbox{\cite{chu2020physical}}} \\ \hline
{HM} & 0.1238 & {\mbox{\cite{chu2020physical}}} \\ \hline
PM and {HM} & 0.0182 & {\mbox{\cite{chu2020physical}}} \\ \hline
\end{tabular}
\label{table1}
\end{table}

\item \textit{Proportion of undetected individuals}. Asymptomatic and pre-symptomatic individuals play an essential role in the spread of the virus.~\cite{moghadas2020implications,li2020substantial} However, the proportion of infected individuals who remain asymptomatic and the transmission rate were not evident during the period when the first simulations were conducted.~\cite{yin2020comparison,he2020temporal} The scientific literature reported different percentages of asymptomatic individuals worldwide.~\cite{Reproduc9:online,wiersinga2020pathophysiology} Based on this information and the testing capacity in the country, we set the percentage of undetected persons at $25\%$. This percentage accounts for asymptomatic individuals and those who were not captured by the health system even with symptoms, either because they did not consult a health care center or due to the country's testing capacities. In mid-2020, Costa Rica conducted 3·3 tests per 10,000 inhabitants on average, with a positivity rate of $25\%$.~\cite{Coronavi67:online, minsaonline} To date, the World Health Organization (WHO) recommendation was that positivity rates should remain at $5\%$ or lower for reopening.~\cite{WHOPosRate}

\item \textit{Protective measures.} Previous studies~\cite{chu2020physical,Cook,Barboza} have shown that incorporating appropriate personal protective measures and restricting human mobility reduce the probability of infection. If a particular interaction between two nodes involves using personal protective measures, then we reduce the probability $\beta_{ij}$ accordingly; see Table~\ref{table1}. We assumed that the percentage of individuals who effectively used personal protective measures in July and August 2020 was $50\%$ (the use of masks was recommended but not mandatory). We also included a parameter that accounts for the percentage of individuals circulating and its impact on human {mobility} (represented as {HM} in Table~\ref{table1}), which is affected by public health restrictions and the consequential social behavior changes. To simulate the different levels of restrictions or the lifting of measures, we reduce or increase the value of this parameter, respectively. Furthermore, we decrease the number of connections (edges) between nodes in different cantons to simulate the mobility restriction between specific locations.

\item \textit{Epidemiological parameters}. Parameters related to hospitalization and ICUs were provided by the Costa Rican Social Security Fund (CCSS), and parameters related to the transmission are from the literature; {see Table~\mbox{\ref{tb:param2}} and hospitalization parameters are provided by the CCSS (see Table~\mbox{\ref{tb:param1}}).} We have used the exact number of days for transitions between states.
\end{itemize}

\begin{table}[h!]
\centering
\caption{{\bf {Hospitalization parameters.}} Values were provided by CCSS.}
\begin{tabular}{|>{\raggedright\arraybackslash}m{6.5cm}|>{\centering\arraybackslash}m{3cm}|>{\centering\arraybackslash}m{2.5cm}|>{\centering\arraybackslash}m{2.5cm}|}
\hline
\rowcolor{gray!10}
Parameter & July 12-16, 2020 & July 22, 2020 & August, 2020  \\ \hline
Diagnosed people requiring hospitalization (\%)& 6.0  & 8.0 & 8.0 \\ \hline
Hospitalized people requiring ICU (\%) & 17 & 17& 18 \\ \hline
Average days a person stays in ICU & 17.4  & 17.4  & 18.1 \\ \hline
Average days a person stays in ward & 9.6  & 10.2  & 10.2\\ \hline
Mortality in ward (\%)& 2 & 4 & 5 \\ \hline
Mortality in ICU (\%)& 60 & 46  & 60 \\ \hline
\end{tabular}
\label{tb:param1}
\end{table}

\begin{table}[!ht]
\centering
\caption{{\bf Epidemiological parameters.} Values are updated throughout the pandemic as new information is available. The probability of infection was estimated using an Approximate Bayesian Computation~\cite{Scott} method at the beginning of the pandemic.}
\begin{tabular}{|>{\raggedright\arraybackslash}m{9.9cm}|>{\centering\arraybackslash}m{1.1cm}|>{\centering\arraybackslash}m{1.1cm}|}
\hline
\rowcolor{gray!10}
Parameter& Value   & Ref. \\ \hline
Average days from onset of symptoms to hospitalization& 7& \cite{liu2020predicting} \\ \hline
Average days from onset of symptoms to ICU& \multicolumn{1}{c|}{11}& \cite{wiersinga2020pathophysiology, CDCPar} \\ \hline
Recovery time (days)& 14& \cite{ScienceNews}\\ \hline
Percentage of diagnosed people& 75\%& \cite{Reproduc9:online,wiersinga2020pathophysiology}\\ \hline
Percentage of diagnosed people who do not isolate themselves & \multicolumn{1}{c|}{10\%}& \cite{yin2020comparison} \\ \hline
Incubation period (days)& 6& \cite{Lauer2020, wiersinga2020pathophysiology} \\ \hline
Probability of infection $\beta_{ij}$& 0.21& Est.  \\ \hline
Percentage of undiagnosed people requiring hospitalization   & 1\%& Est.  \\ \hline
Mortality out of hospitalization & 0.008\% & Est. \\ \hline
\end{tabular}
\label{tb:param2}
\end{table}

\subsection*{Scenarios performed in July 2020}

{In response to the requests that arose from the health authorities during different periods, we proposed three scenarios to evaluate and observe the impact of interventions on reducing cases and hospital admissions. Each scenario was designed and implemented on different dates. Differences were related to the reopening assumption after the ``epidemiological fence" and the percentage of people adhering to the restrictions.}



\begin{enumerate}
    \item Scenario 1: July 12, 2020. 
    \begin{itemize}
        \item[--] Restrictions based on canton alert status are removed from the simulation after July 19, 2020.
        \item[--] It is assumed that the $70\%$ of the country maintained human mobility during that period.
    \end{itemize}
    \item Scenario 2: July 16, 2020.
        \begin{itemize}
        \item[--] Restrictions based on canton alert status are maintained until August 2, 2020 only in cantons that remain in orange alert status.
         \item[--] It is assumed that the $70\%$ of the country maintained human mobility during that period.
    \end{itemize}
    \item Scenario 3: Jul 22, 2020.
        \begin{itemize}
        \item[--] Restrictions based on canton alert status are removed from the simulation after July 19, 2020.
        \item[--] Percentage of individuals that maintained human mobility changed gradually in places with an orange alert status: $70\%$ from July 11 to 19, 2020, $60\%$ from July 20 to 31, 2020, and $50\%$ after July 31, 2020. In cantons with a yellow alert status, $50\%$ of the population maintained human mobility.
    \end{itemize}
\end{enumerate}

{In all the scenarios, we assumed that (i) $50\%$ of the population used protective measures and the virus transmission probability is reduced accordingly, and (ii) the number of contacts is sampled from a uniform distribution; see Table~\mbox{\ref{table1}}. We also assumed that from July 11-19, 2020, individuals living in areas where the ``epidemiological fence" was implemented could only contact people from areas with the same alert status.}\\

\subsection*{Scenarios performed in August 2020}

In August 2020, public health authorities {decided to implement a strategy} of closing and opening periods called the {\it hammer and dance};~\cite{pueyo2020coronavirus} details about the interventions are on the official website of the Ministry of Health of Costa Rica.~\cite{minsaonline}

{To help inform the potential effects of this new strategy, we simulated a set of scenarios considering several closing periods of 12 days and opening periods of nine days for two and a half months.} Human mobility restriction was simulated considering two assumptions: the first related to the percentages of individuals who contributed to new infections (human mobility parameter ({HM})). The canton alert differentiated the parameter value. The second was related to the connections of the individuals (nodes) with people of different cantons (mobility restriction (MR)). We restricted the interaction of individuals in cantons in orange alert to only contact individuals within the same canton. On the other hand, people living in cantons with a yellow alert could contact those with the same yellow alert. Scenarios are summarized in Table~\ref{tab:AugustScenarios}.


\begin{table}[!ht]
\centering
\caption{Scenario assumptions performed in August, 2020. MR stands for mobility restrictions between cantons with different levels of alert. In the model, people living in cantons with an orange alert only had interactions with people living in the same canton.}
\resizebox{\columnwidth}{!}{
\begin{tabular}{|p{2cm}|p{2cm}|p{4cm}|p{4cm}|p{4cm}|p{4cm}|p{4cm}|}
\hline
\rowcolor{gray!10}
\centering Level of alert & \centering Open/Close strategy&
\begin{minipage}[t]{\linewidth}%
\centering \textbf{Scenario 1 50\%-70\%} \end{minipage}
&
\begin{minipage}[t]{\linewidth}%
\centering \textbf{Scenario 2 50\%-60\%} \end{minipage}
&
\begin{minipage}[t]{\linewidth}%
\centering \textbf{Scenario 3 50\%-65\%} \end{minipage}
&
\begin{minipage}[t]{\linewidth}%
\centering \textbf{Scenario 4 45\%-55\%} \end{minipage}
&
\begin{minipage}[t]{\linewidth}%
\centering \textbf{Scenario 3 30\%-40\%} \end{minipage}\\
\hline
\hfil Yellow & Dance & 
\centering 50\% {HM}&
\centering 50\% {HM}& 
\centering 50\% {HM}&
\centering 45\% {HM} & \multicolumn{1}{c|}{30\% {HM}}\\
\hline
\hfil Yellow & Hammer & 
\centering 50\% {HM} &
\centering 50\% {HM} &
\centering 50\% {HM} & \centering 45\% {HM} & \multicolumn{1}{c|}{30\% {HM}}\\
\hline
\hfil Orange & Dance & 
\begin{minipage}[t]{1\linewidth}%
\centering \textbf{9 days open}\\
\centering 50\% {HM}
\end{minipage}
&
\begin{minipage}[t]{1\linewidth}%
\centering \textbf{9 days open}\\
\centering 50\% {HM}
\end{minipage}
&
\begin{minipage}[t]{1\linewidth}%
\centering \textbf{9 days open}\\
\centering 50\% {HM}
\end{minipage}
&
\begin{minipage}[t]{\linewidth}%
\centering\textbf{9 days open} \\45\% {HM} 
\end{minipage}
&
\begin{minipage}[t]{\linewidth}%
\centering \textbf{9 days open}\\30\% {HM} 
\end{minipage}\\
\hline
\centering Orange &Hammer & \begin{minipage}[t]{\linewidth}%
\centering \textbf{12 days closed}\\70\% {HM} and MR
\end{minipage}
&
\begin{minipage}[t]{\linewidth}%
\centering \textbf{12 days closed}\\60\% {HM} and MR
\end{minipage}
&
\begin{minipage}[t]{\linewidth}%
\centering \textbf{12 days closed}\\65\% {HM} and MR
\end{minipage}
&
\begin{minipage}[t]{\linewidth}%
\centering \textbf{12 days closed}\\ 55\% {HM} and MR
\end{minipage}
&
\begin{minipage}[t]{\linewidth}%
\centering \textbf{12 days closed}\\40\% {HM} and MR
\end{minipage}\\ 
\hline
\end{tabular}%
}
\label{tab:AugustScenarios}
\end{table}

During the hammer period, restrictions on vehicular circulation and the opening of commerce were applied only in the cantons with an orange alert status. Every week, the alert status changed. As an attempt to predict which cantons would change status, we introduced an indicator given by:

\begin{equation*}
\text{{Cantonal Hazard Rate}} = \dfrac{\dfrac{\text{new cases in the last 3 weeks in a canton}}{\text{canton population}}}{\dfrac{\text{new cases in the last 3 weeks in the country}}{\text{country population}}}
\end{equation*}
{If the Cantonal Hazard Rate is greater than one in a canton with a yellow alert status, it changes to an orange alert.} All cantons with an orange alert status do not change status. Finally, connections between nodes are only allowed in places with the same alert status (yellow or orange) in the hammer phase.

\section*{Results}

All simulations were implemented in Matlab R2020a~\cite{Matlab} and run in two remote servers: (i) a Dell PowerEdge R740 with 64GB of RAM and two Intel®(R) Xeon®(R) Silver 4114 CPU @ 2.20GHz processors, and (ii) a Lenovo SR650, with two Intel® Xeon® Plata 4214, 2.20 GHz processors with 128GB of RAM. The results presented below correspond to the simulations and scenarios implemented in July and August 2020, modeled in real-time. Therefore, the number of simulations per scenario varies between 20 to 60 since it depends on the response time. We present the average of the total simulations in each scenario.

\subsection*{Scenarios conducted in July 2020}

\begin{figure}[!ht]
\centering
\includegraphics[scale=0.3]{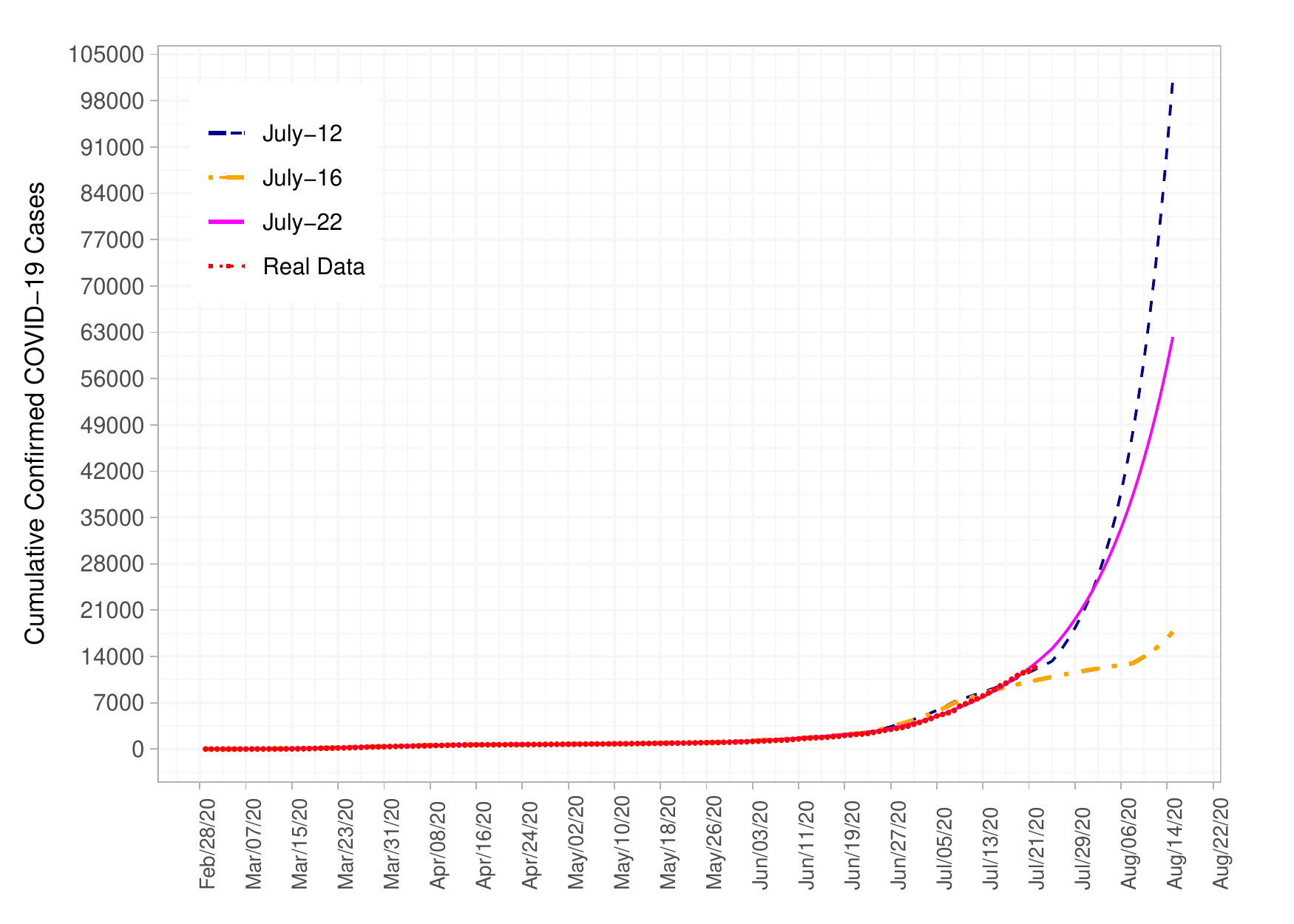}
\includegraphics[scale=0.3]{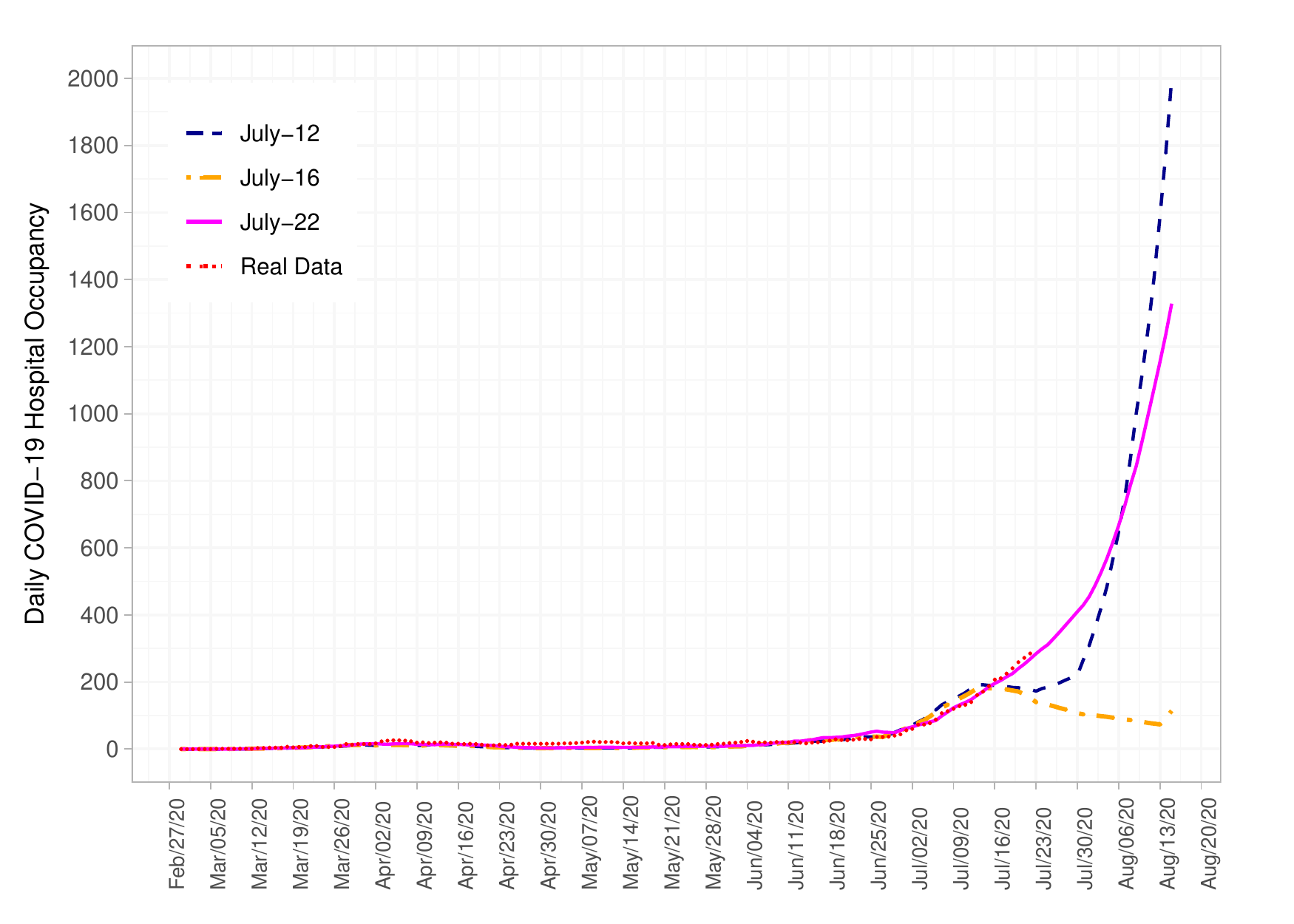}
\includegraphics[scale=0.3]{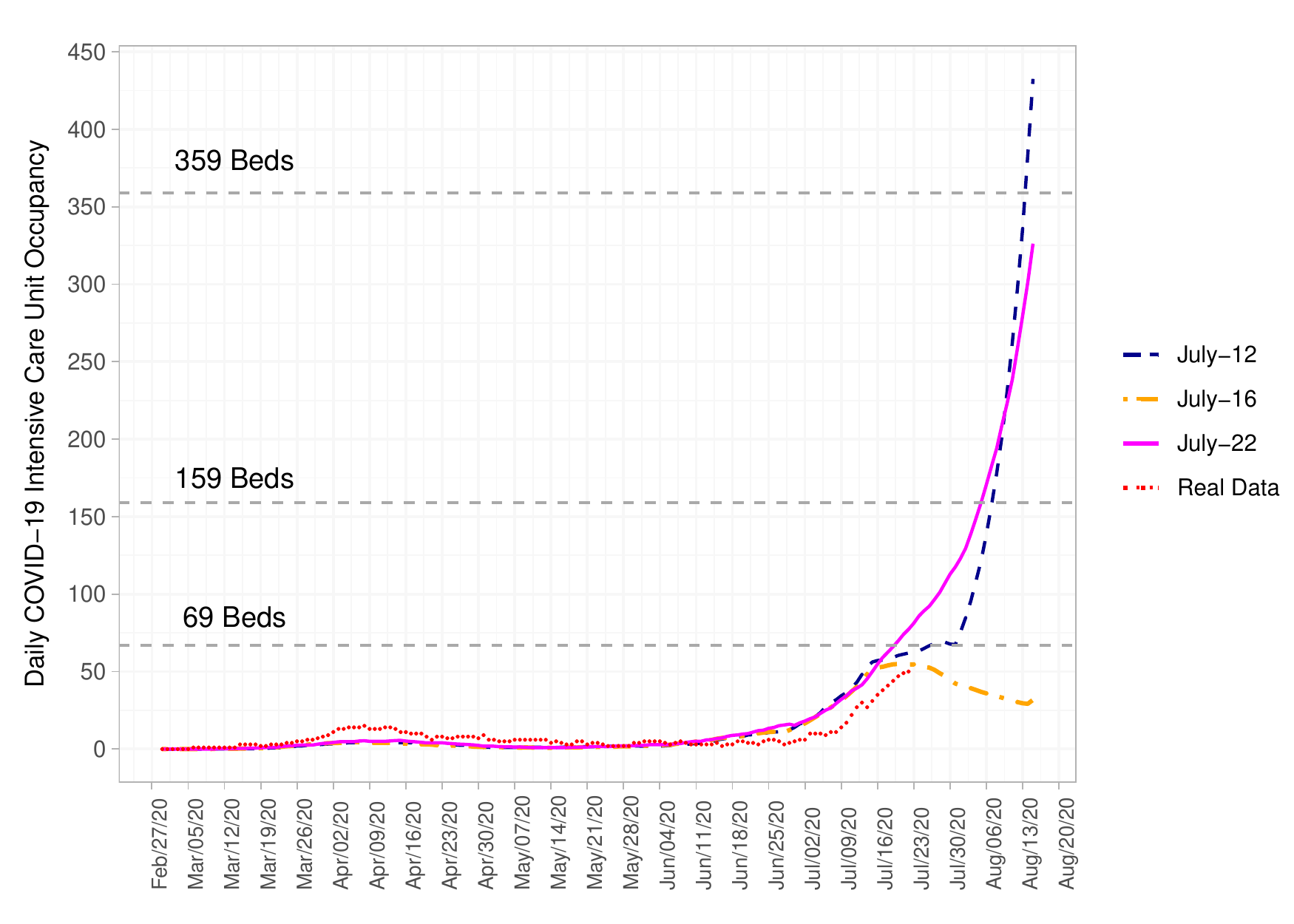}
\caption{{\bf Scenario in July:} {Cumulative confirmed Covid-19 cases (Left), hospitalizations (Center), and ICU (Right) admissions.}}
\label{fig:JulySce1}
\end{figure}

Our results showed that infections and hospitalizations were expected to increase abruptly if, after only a week of restrictions, measures were lifted to levels similar to those before the epidemiological fence (July 12, 2020, scenario). This rapid growth could have been controlled partially if the reopening occurred gradually, as shown in the July 22, 2020, scenario. However, if restrictions were maintained for three weeks in cantons under orange alert (July 16, 2020, scenario), the model showed a significant reduction in infections, hospitalizations, and ICU admissions; see Fig~\ref{fig:JulySce1}. 

\subsection*{Scenarios conducted in August 2020}

Results showed that establishing the three closing phases of the ``Hammer and Dance" period was expected to curtail the rise in cases successfully, but only if they could restrict population mobility in ranges over $50\%$. Furthermore, ICU admissions were only expected to remain within the 159 threshold with $50-65\%$ population adherence.\\

\begin{figure}[!ht]
\centering
\includegraphics[scale=0.3]{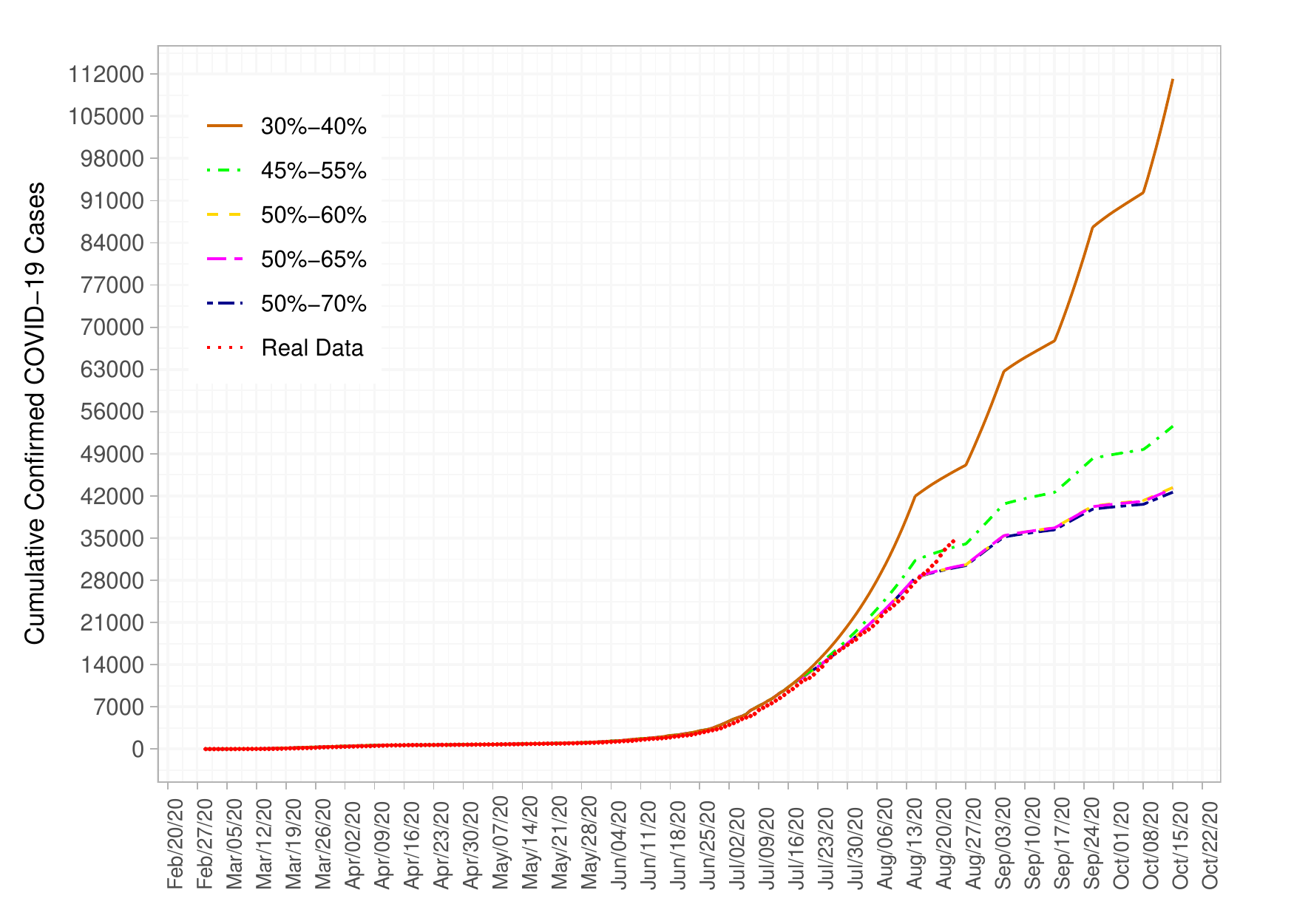}
\includegraphics[scale=0.3]{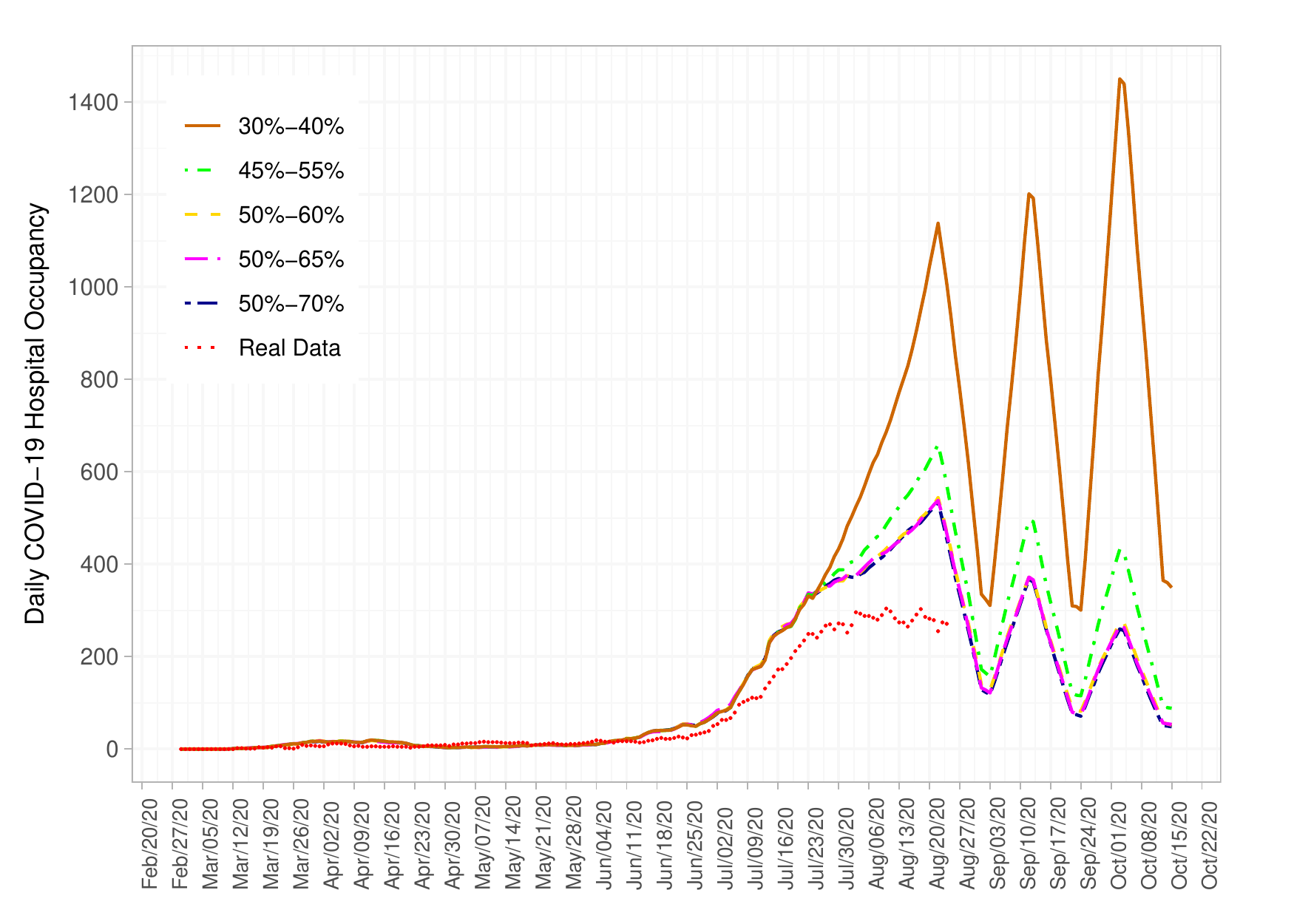}
\includegraphics[scale=0.3]{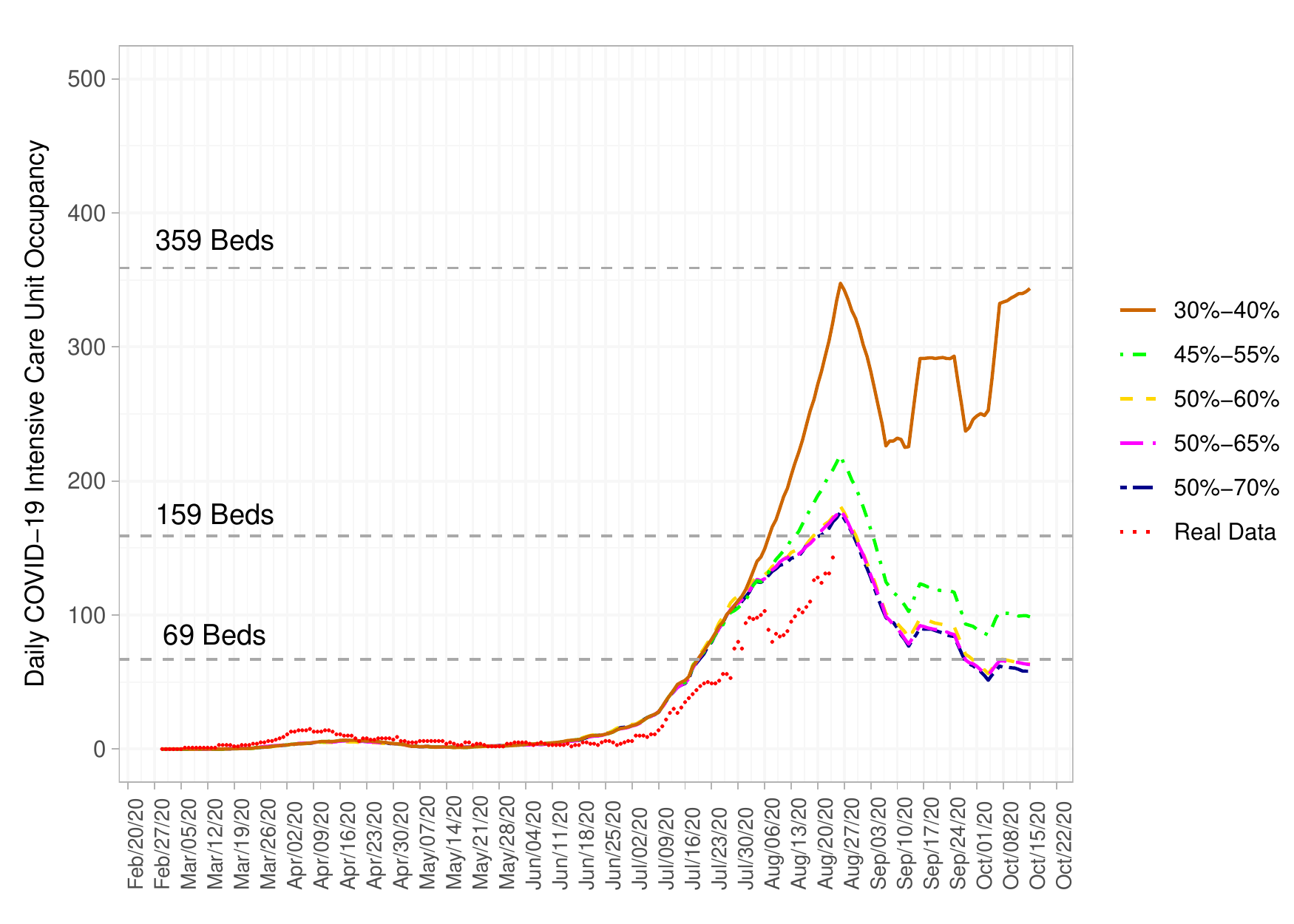}
\caption{{\bf Scenario in August:} {Cumulative confirmed Covid-19 cases (Left), hospitalizations (Center), and ICU (Right) admissions. Percentages in the legend correspond to changes in the human mobility (HM) parameter. The first values are for cantons with a yellow alert, and the second values are for cantons with an orange alert.}} 
\label{fig:ScenariosAugust1}
\end{figure}

There were also three abrupt decreases in hospitalization as a consequence of the three closing periods and the assumptions related to the connectivity of individuals since we assumed that each node only interacted with contacts in cantons with the same yellow alert and restricted the connection to people in the same canton for those living in cantons with orange alert see Fig~\ref{fig:ScenariosAugust1}.

\section*{Discussion} 

A multilayer network model was developed in a joint effort between members from the Ministry of Health, the Pan American Health Organization (PAHO), and scientists from academia. The model's flexibility allowed us to incorporate a socio-demographic component, public health interventions, and the basic epidemiological characteristics of the virus transmission. The primary use of the model focused on three aspects: short and mid-term projection of the number of cases and hospital demand, simulation of hypothetical scenarios that helped health authorities and government officials in the design of preventive control measuresVaccine administrated by locationVaccine administrated by location, and finally, on simulating the impact of interventions (results were reported by PAHO in the monthly reports on the Covid-19 pandemic in Costa Rica).~\cite{PAHO} \\

The findings presented here show the experience of real-time modelling scenarios developed according to the necessities of Costa Rican public health officials during July and August 2020. After lifting most restrictions due to a low case count, the country started experiencing increased cases and hospitalizations. This increase led decision-makers to implement an ``epidemiological fence" from 11 to July 19, 2020. Scenarios projected on July 12, 2020, warned of an expected abrupt increase in cases and hospitalizations if restrictions were lifted after only a week. Instead, maintaining them for three weeks was expected to contain the growth in disease transmission. \\

In August 2020, decision-makers needed to balance public health with economic recovery. The Hammer and Dance Period was an attempt to achieve precisely that. Results showed that such strategy was expected to contain the disease only if population mobility was reduced in ranges over $50\%$ during closure phases. We did not know for sure if and how effective these sanitary measures were at reducing human mobility. We performed a parallel analysis~\cite{Barboza} describing the impact that reduction in mobility had on infections and hospitalization. However, we also anticipated that the same intervention imposed at different pandemic stages was unlikely to produce the same results. A population growing tired of confinement and those with precarious employment are likely to countermeasure the effect of restrictions. The use of masks was not mandatory at that time but was under consideration. In our model, we set the use of personal protection measures at $50\%$. This parameter became a factor of adjustment for subsequent scenarios.\\

{It is crucial to highlight the implications of the analysis in hospital occupancy. By July 2020, Costa Rica had been working on increasing its ICU capacity from 67 to 159 beds (0.013 to 0.03 beds per 1000 hab) for Covid-19 and had the goal to reach 359 (0.07 beds per 1000 hab) by the end of the year. Intensive care services are offered mainly by the three leading hospitals in the country, the Calder\'on Guardia Hospital, the Mexico Hospital, and the San Juan de Dios Hospital located in the city of San Jos\'e also known as the Great Metropolitan Area. Therefore, patients must be transferred from anywhere in the country to the Great Metropolitan Area, which implies additional logistical considerations.} Lifting restrictions after only one week of epidemiological fence presented a scenario of rapid collapse in hospital capacity. Our results suggested that adopting a Hammer and Dance strategy could potentially contain this collapse and gain valuable time for the health care system to reach its goal of 359 ICU beds. However, these measures alone could not prevent this collapse if population adherence to mobility restrictions was lower than $55\%$. Furthermore, with adherence lower than $45\%$ the projected ICU admissions could surpass the threshold of 359 ICU beds. After the public health interventions, the average ICU occupancy in July and August was 0.0068 and 0.0022 patients per 1000 inhabitants, respectively. \\

One of the challenges and limitations of the analysis is how difficult it is to know what variable is attributable to which effect. Sanitary measures were imposed in an ever-changing social scenario where people's adherence is as relevant as the interventions themselves. If our model predicted that certain restrictions, based on past data, were going to reduce cases and then cases did not drop, it is as likely that the people responded differently to such measures as that the model was wrong. Indeed, it would probably be a combination of both, plus several other factors. Hence, the model's value must not be judged by the accuracy of its forecasts but by the value in informing decisions and their impact in containing the disease, the collapse of health care systems, and ultimately saving lives, highlighting the value of close collaboration between researchers and policy-makers.\\ 

The ongoing global health crisis has evidenced the challenges of modeling a real-time pandemic and the importance of having an interdisciplinary research team working hand in hand with health authorities. Effective communication between scientists, public health entities, and government officials is vital in developing mathematical, statistical, and computational models that can provide insight to create successful disease prevention and control tools in a practical setting.\\

Counting with adequate, robust models, sophisticated enough to capture the changing dynamics and variety of factors interacting simultaneously in a pandemic of the scale of Covid-19 could be limited by the computational capacity and the available data. Furthermore, what we know about the virus is still very limited. After more than a year of this pandemic, virus transmission dynamics are still a subject of research. Many open questions remain, and the assumptions have to be constantly updated.


\section*{Contributors}
Y.G. and P.V. designed the study, performed simulations, analyzed data, and wrote the initial draft. G.M. and F.S. designed the study, interpretation and analysis of the results, and edited the manuscript. J.C. coded the model and edited the manuscript. L.B., T.R. and M.D.P. designed the study and edited the manuscript. All authors reviewed the manuscript. G.M. and M.D.P. are staff members of the Pan American Health Organization. The authors alone are responsible for the views expressed in this publication, and they do not necessarily represent the decisions or policies of the Pan American Health Organization.  

\section*{Acknowledgments}
The authors would like to thank the Research Center in Pure and Applied Mathematics and the School of Mathematics at Universidad de Costa Rica for their support during the preparation of this manuscript. They also thank the Ministry of Health and CCSS for providing data and valuable information for this study.

\nolinenumbers

\end{document}